\documentclass[12pt]{iopart}

\usepackage[english]{babel}
\usepackage{amssymb}
\usepackage{epsfig}
\def\lsim{\raisebox{-.4ex}{$\stackrel{<}{\scriptstyle \sim}$\,}}

\begin{document}
\newcommand{\Od}{{\cal O}}
\newcommand{\degree}{^\circ}
\newcommand{\K}{\textrm{K}}


\title{Large-scale cosmic flows and moving dark energy}%

\author{Jose Beltr\'an Jim\'enez and Antonio L. Maroto
}
\address{Departamento de  F\'{\i}sica Te\'orica I,
Universidad Complutense de Madrid, 28040 Madrid, Spain.}
\date{\today}

\begin{abstract}
Large-scale matter bulk flows with respect to the cosmic microwave background
 have very recently been detected on scales $\lsim 100h^{-1}$ Mpc  and
$\lsim 300h^{-1}$ Mpc by using  two
different techniques showing an excellent  agreement in the
 motion direction. However, the unexpectedly large measured amplitudes are 
difficult to  understand within the context of standard $\Lambda$CDM
cosmology.  In this work we show
that the existence of such a flow could be signaling the
presence of  moving dark energy at the time when
photons decoupled from matter. We also comment on the relation between the
direction of the CMB dipole and the preferred axis observed in the
quadrupole in this scenario.

\end{abstract}

\maketitle 

\section{Introduction}

Standard Model of cosmology describes an isotropic and homogeneous
universe on very large scales which contains (dark and baryonic)
matter, radiation and dark energy. However, at smaller scales, 
the universe is no longer homogeneous since matter density perturbations 
tend to form structures 
that originated by  primordial inflationary
fluctuations. Thus,  small volumes of matter would acquire 
peculiar velocities with
respect to the CMB rest frame. These peculiar velocities are due to
the gravitational potential of  outer structures and depend
crucially on the matter power spectrum. In standard $\Lambda$CDM, the
large scale matter rest frame coincides with the CMB rest frame so
that as we take larger and larger volumes of matter, peculiar
velocities due to statistical fluctuations should become smaller and smaller.
In fact, peculiar velocities are a very powerful tool in order to probe 
matter density fluctuations on very large scales and for
that reason a great effort has been made 
in order to measure them by using different tracers:
galaxies \cite{galaxies}, clusters \cite{clusters} or Type Ia
supernovae \cite{SNIa}. Although on small scales
($\lsim 60h^{-1}$ Mpc)
 surveys based on
different distance indicators seem to agree with
predictions of $\Lambda$CDM cosmology,  on larger scales
they have seemed to be in conflict
among them for many years, yielding
peculiar velocities in the wide range $0-1000$ km/s. 
However, new data analysis performed in 
recent years \cite{recent} suggest that most of the surveys could also agree
with each other on such large scales. 

A completely new approach to obtain peculiar
velocities of large volumes of matter which makes use of 
the Kinematic
Sunyaev-Zeldovich effect on the CMB photons by the hot gas in
clusters of galaxies was proposed in \cite{KA-B2000} and carried
out very recently in \cite{atrio}. They find coherent bulk flows
on scales of $300h^{-1}$ Mpc towards $l=283\degree\pm\; 14\degree$,
$b=11\degree\pm\; 14\degree$ (in galactic coordinates). 
The reported amplitude for those
peculiar velocities are in the range 600-1000 km/s, although the
authors point out that, even though there is no doubt about the
existence and direction of the flows, the obtained values for the
amplitudes may have some systematic offset. 
In \cite{atrio}, they attribute these
peculiar velocities to pre-inflationary super-Hubble
perturbations (see \cite{turner}).

Also, in a very recent work \cite{Feldman}, a calculation of peculiar
velocities using some of the available measurements has been
performed, but with a new method which allows to reduce the
sensitivity to small scale power and makes possible to compare the
results obtained from different surveys. In that work, they
find a consistent flow of matter on scales of 100$h^{-1}$ Mpc
towards $l=287\degree\pm\; 9\degree$, $b=8\degree\pm\; 6\degree$
 and with an amplitude of $407\pm 81$
km/s. The direction of the 
detected flow  is in very good
agreement with the results in \cite{atrio}, in spite of having
used a very different method. 
The authors of this work claim that these peculiar
velocities may be due to structures larger than the reached scale
of 100$h^{-1}$ Mpc.

In conclusion, these recent results suggest the existence of a
coherent flow of
matter with respect to the CMB rest frame on very large scales. In
both works, such a flow is explained resorting to very large-scale
matter perturbations, i.e., the observed peculiar velocities would be 
caused by the existence of some super-structure which must be further
away than the Great Attractor, located at $40-60 h^{-1}$ Mpc from
us \cite{GA}. In any case, the presence of flows with such a large
amplitude seems to be difficult to understand in the
context of standard $\Lambda$CDM cosmology, which
predicts much smaller velocities on the quoted scales.

In this work, we propose an alternative explanation
in which the observed bulk flow could be indicating that
 matter is globally moving with respect to radiation,
i.e., that matter and radiation do not share a common rest frame
on very large scales.
In other words, the velocity of a matter
bulk of size $R$ may have two independent components:
$\vec{V}_{R,bulk}=\vec{V}_{R,sta}+\vec{V}_{cosmic}$ where
$\vec{V}_{R,sta}$ is the statistical r.m.s. velocity fluctuation
generated by density inhomogeneities and $\vec{V}_{cosmic}$ is
the cosmic velocity of
matter with respect to the CMB because of having a different rest
frame. Then, as we average over a large volume of matter, 
the first term becomes negligible whereas the second
term remains constant and gives rise to a net cosmic flow.
However, if that is the actual situation, then the question arises as
how matter can be currently moving with respect to photons given the fact that 
they were  strongly coupled in the early universe.
In this work we propose that such velocity offset can be generated
by the motion of dark
energy with respect to the primordial plasma in the early universe.

\section{Moving dark energy}

In previous works \cite{dipole1,quadrupole}, the
possibility of having dark energy with a different large-scale
rest frame from that of the matter-radiation plasma in the early
universe has been studied. If dark energy is a perfect fluid 
which has always been
decoupled from matter and radiation, there is no reason to expect
it to have the same rest frame as the other of components of the
universe. In such a case, the relative dark energy velocity $v_{DE}$
should be considered as a free cosmological parameter, on
equal footing with the density parameter $\Omega_{DE}$ or
equation of state $w_{DE}$, and accordingly it could only be
determined from observations.
Indeed, the fact of having moving dark energy at the
time when photons decoupled from matter allows to have relative
velocities between dark matter, baryons and photons. This relative
motion may be precisely the one detected in \cite{atrio,Feldman}.

The model proposed in \cite{dipole1,quadrupole} consists
of a universe filled with four perfect fluids: baryons, radiation,
dark matter and dark energy, whose velocities can differ from each
other. In that case, we can define the velocity of the
Cosmic Center of Mass (CCM), for small velocities, as:
\begin{eqnarray}
\vec{S}=\frac{\sum_\alpha(\rho_\alpha+p_\alpha)\vec{v}_{\alpha}}
{\sum_\alpha(\rho_\alpha+p_\alpha)},
\end{eqnarray}
where $\alpha$ stands for the four aforementioned components of
the universe. Now, if we place ourselves in the CCM rest frame,
defined by the condition $\vec{S}=0$, the following relation must
hold:
\begin{equation}
\sum_\alpha(\rho_\alpha+p_\alpha)\vec{v}_{\alpha}=0.\label{CCM}
\end{equation}
In the early universe, radiation dominates over the rest of
components so it will drag baryons and dark matter particles in
such a way that they all will share a common rest frame, as 
expected for interacting species in thermal equilibrium.
Concerning dark energy, since it does not interact with
photons, it can move in a different way. However, as long as we
use the CCM rest frame, photons and dark energy velocities are
related by means of (\ref{CCM}) as follows:
\begin{equation}
\vec{v}_{DE}^{early}=-\frac{4}{3}\frac{\rho^{early}_R}
{\rho^{early}_{DE}+p^{early}_{DE}}\vec{v}^{early}_R\label{velrel}
\end{equation}
where we have used that matter and radiation velocities are the
same and that matter energy density is negligible compared to that
of radiation. Notice that this relation provides a preferred axis
in the universe given by the fluids motion direction. 
In other words, the universe will have axial symmetry
around the axis defined by the motion of the different components.

On the other hand, the momentum conservation equation for each
fluid reads:
\begin{equation}
\frac{d}{dt}\left[a^4(\rho_\alpha+p_\alpha)\left(\vec{S}
-\vec{v}_\alpha\right)\right]=0\label{cons}
\end{equation}
which can be immediately integrated assuming constant equation of
state $w_\alpha=\frac{p_\alpha}{\rho_\alpha}$ to give:
\begin{equation}
\vec{S}-\vec{v}_\alpha=\vec{v}_\alpha^0a^{3w_\alpha-1}\label{velevol}
\end{equation}
with $\vec{v}_\alpha^0$ the present value of the velocity. Thus,
radiation ($w_R=\frac{1}{3}$) moves with constant velocity with
respect to the CCM, whereas matter ($w=0$) velocity decays as
$a^{-1}$. The evolution of dark energy velocity depends on the
particular model under consideration. Therefore, according to
(\ref{velevol}), matter and radiation have constant velocity before
decoupling, but, after that, matter starts reducing its velocity
as the universe expands whereas radiation keeps moving with the same
constant velocity. As a consequence, the presence of moving dark 
energy at the
time of recombination makes possible that matter and radiation
could acquire a relative velocity after decoupling.
This effect can only take place if dark energy is not a
cosmological constant, otherwise there would be no momentum contribution
from dark energy to (\ref{CCM}).
Notice that, since dark matter particles should
decouple before baryons do (in order to be able to form the structures we
observe today) and they both have the same equation of
state, we expect them to have a relative motion with
constant velocity. Thus, the complete picture of the motions would be
as follows: 
radiation moving with constant velocity, baryons and dark matter
moving in the same direction with constant relative velocity and,
both, slowing down with respect to photons. Finally, dark energy
would move along the opposite direction.

\section{CMB dipole from moving dark energy}

In the CCM rest frame, the metric becomes diagonal and
can be written as that of an axisymmetric Bianchi I space-time. 
Moreover,
as the velocities of the fluids are small, this metric can be
treated as a Robertson-Walker metric with  a small
perturbation. Thus, the usual expression for the dipole given by the
Sachs-Wolfe effect at first order applies to this case, although
we have to read the velocities as relative to the CCM.
Hence, the dipole in an arbitrary frame for this model is given by
\cite{dipole1}:
\begin{equation}
\frac{\delta
T_{dipole}}{T_0}\simeq\vec{n}\cdot\left(\vec{S}-\vec{v}\right)^0_{dec}
\label{dipoleq}
\end{equation}
where $\vec{n}$ is a unitary vector along the direction of
observation, $\vec{v}_0$ is the velocity of the observer today and
$\vec{v}_{dec}$ is the velocity of the emitter at decoupling time.
Notice that according to (\ref{dipoleq})  
when several fluids with relative velocities are present, the
CMB dipole is given by the  velocities of emitter and observer with 
respect to the CCM frame. In the case in which all the fluids share a common
rest frame, the velocites are referred to that frame which is nothing but 
the CMB frame. However, when the fluids velocities are different, 
the physically relevant frame for the dipole is the CCM which is different
from the CMB frame. In this sense, if emission took place
from a source at rest with respect to the CCM frame, that frame could be 
determined physically as  the frame attached to an observer who
measures a vanishing dipole.

If we assume that the intrinsic dipole fluctuation at the last
scattering surface is negligible we can take
$\vec{v}_{dec}\simeq\vec{v}_R^{dec}$. On the other hand, if the
observer today is at rest with respect to matter we have
$\vec{v}_0\simeq\vec{v}_M^0$. Thus, if we now refer the
velocities to the CCM frame, we finally get:
\begin{equation}
\frac{\delta
T_{dipole}}{T_0}\simeq\vec{n}\cdot\left(\vec{v}_R^0-\vec{v}_M^0\right)
\label{dipole}
\end{equation}
where we have used that the velocity of radiation  with respect to the
CCM is constant.
Then, according to (\ref{dipole}), we conclude that the  contribution
to the dipole temperature fluctuation of
the CMB is due to the relative motion of matter with respect to
radiation. This is precisely the kind of flow detected in
\cite{atrio,Feldman} so the direction reported in both papers
gives directly the direction of motion of the fluids.

Notice that, although calculated in the CCM frame, (\ref{dipole})
is valid for any
frame since it is expressed as the difference of two velocities
evaluated at the same time. According to (\ref{velevol}), today the
velocity of matter with respect to the CCM is expected to be much smaller
than that of
radiation  and the dipole can also be written as:
\begin{equation}
\frac{\delta
T_{dipole}}{T_0}\simeq\vec{n}\cdot \vec{v}_R^0\label{dipole2}
\end{equation}
so that the cosmological dipole could  be alternatively interpreted 
as due to the relative motion of radiation with respect to the CCM.

Using again expression (\ref{CCM}), it is possible to relate
the amplitude of the dipole  (${v}_R^0$) to the present
value of the dark energy velocity with respect to the CCM frame:
\begin{eqnarray}
v_{DE}^0=\frac{v_R^0}{(1+w_{DE}^0)\Omega_{DE}}\left(\frac{2}{3}\Omega_R
+\frac{\Omega_B}{1+z_{dec}}+\frac{\Omega_{DM}}{1+z_{*}}
\right)
\end{eqnarray}
where $z_*$ is the decoupling redshift of dark matter and
radiation and $w_{DE}^0$ is the present value of the dark energy
equation of state. We can estimate typical values of the present
dark energy velocity. Thus, from the measured bulk flows we can
take $v_R^0\sim 500$ km/s and assuming $\Omega_{DE}\simeq 0.7$, $z_*>
10^{5}$ and $w_{DE}^0\simeq -0.97$, we get $v_{DE}^0\sim 1$ km/s.

\section{Effects on the quadrupole}

The relative motion of the different components in the universe
also affects higher multipoles of the CMB. In general, the
contribution to the $\ell$th multipole  will be $\sim
|\vec{v}|^{\ell}$ so the effect decreases rapidly as we increase
$\ell$ because of the smallness of the velocities.

In particular, for quadrupolar temperature fluctuation it is possible
to obtain \cite{quadrupole}:
\begin{equation}
\frac{\delta
T_Q}{T_0}=-\frac{1}{2}h\left(\hat{v}_i\hat{v}_j-\frac{1}{3}\delta_{ij}\right)n^in^j,
\end{equation}
where $h$ is a function which measures the variation of the degree
of anisotropy from the last scattering surface until present
and depends on the particular dark energy model under
consideration and $\hat{v}_i$ is a unitary vector pointing along
the direction of the motion of the fluids. Moreover, if such a
direction is given by $(\theta,\phi)$, the components of the
quadrupole are:
\begin{eqnarray}
&a_{20}^A&=\frac{\sqrt{\pi}}{6\sqrt{5}}h[1+3\cos2\theta],\nonumber \\
&a_{21}^A&=-(a_{2-1}^A)^*=-\sqrt{\frac{\pi}{30}}h\,e^{-i\phi}
\sin 2\theta ,\nonumber \\
&a_{22}^A&=(a_{2-2}^A)^*=\sqrt{\frac{\pi}{30}}h\,e^{-2i\phi}
\sin^2\theta .
\end{eqnarray}
Then, the quadrupole generated by the motion is:
\begin{equation}
Q_A\equiv\sqrt{\frac{3}{5\pi}\sum_{m=-2}^{2}\left|a_{2m}^A\right|^2}=\frac{2}{5\sqrt{3}}h.
\end{equation}
On the other hand, we still have to add the isotropic quadrupole
$Q_I$ coming from the inflationary fluctuations, whose components
can be written as:
\begin{eqnarray}
&a_{20}^I&=\sqrt{\frac{\pi}{3}}\,e^{i\alpha_1}Q_I,\nonumber\\
&a_{21}^I&=-(a_{2-1}^I)^*=\sqrt{\frac{\pi}{3}}\,e^{i\alpha_2}Q_I,\nonumber\\
&a_{22}^I&=(a_{2-2}^I)^*=\sqrt{\frac{\pi}{3}}\,e^{i\alpha_3}Q_I\label{QI},
\end{eqnarray}
where the phase factors $\alpha_i$ can be considered as stochastic
variables. As the anisotropy is expected to be small, the total
effect is given by the linear superposition of both contributions
\cite{Bunn}, i.e., $a^T_{\ell m}=a^A_{\ell m}+a^I_{\ell m}$. Now,
in a frame with the $z$-axis parallel to $\hat{v}_i$ we have
$\theta=0$ so that the only non-vanishing anisotropic component is
$a_{20}^A$. That way, the only modified component of the total
quadrupole, in that frame, is $a_{20}$.  However, in a
general frame with the velocities along the direction
$(\theta,\phi)$ the total quadrupole becomes \cite{Cea}:
\begin{equation}
Q^2_T=Q^2_A+Q_I^2-2fQ_AQ_I
\end{equation}
where
\begin{eqnarray}
f&=&\frac{1}{4\sqrt{5}}\left[2\sqrt{6}\left[-\sin\theta\cos(2\phi
+\alpha_3)\right.\right. \\
&+&\left.\left. 2\cos\theta\cos(\phi+\alpha_2)\right]\sin\theta
-(1+3\cos(2\theta))\cos\alpha_1\right]\nonumber.
\end{eqnarray}
Assuming that the anisotropic quadrupole is a new contribution to
add on top of the true model which provides the observed
quadrupole, it has to satisfy \cite{quadrupole}:
\begin{eqnarray}
0\;\mu\K^2\lsim &(\delta T_A)^2& \lsim 1861\;\mu\K^2 \;\;\ 68\% \;\mbox {C.L.}\nonumber \\
0 \;\mu\K^2\lsim &(\delta T_A)^2& \lsim 5909\;\mu\K^2 \;\;\ 95\%
\;\mbox {C.L.} \label{excluded}
\end{eqnarray}
in order not to have a too large total quadrupole. These
conditions yield the following constraints on the parameter $h$:
\begin{eqnarray}
0\;\lsim &h& \lsim\; 6.92\times10^{-5} \;\;\ 68\% \;\mbox {C.L.}\nonumber \\
0\;\lsim &h& \lsim\; 1.23\times10^{-4} \;\;\ 95\% \;\mbox {C.L.}
\label{hexcluded}
\end{eqnarray}

\section{Model example: scaling dark energy}

Scaling models \cite{scaling} are those with equation of state
such that dark energy mimics the dominant component of the
universe throughout most of the universe evolution. Thus, dark
energy evolves as radiation before matter-radiation equality and
as matter after that. However, in order to explain the accelerated
expansion of the universe, dark energy has to exit from that
regime and join into one with $w_{DE}<-1/3$ at some point. Then,
the evolution of the dark energy density is given by:
\begin{eqnarray}
\rho_{DE}=\left\{\begin{array}{c}
\rho_{DE}^0\;a_{T}^{-3w_{DE}}a_{eq}a^{-4}
\;\;\;\;\;\;\;\;\;\;\;\;\;a<a_{eq}\\\\\\
\rho_{DE}^0\;a_{T}^{-3w_{DE}}a^{-3}
\;\;\;\;\;\;\;\;\;a_{eq}<a<a_{T}\\\\\\
\rho_{DE}^0\;a^{-3(w_{DE}+1)}
\;\;\;\;\;\;\;\;\;\;\;\;\;\;\;\;\;a>a_{T}\end{array} \right.
\end{eqnarray}
where as commented before, $a_{T}$ is the scale factor when
dark energy leaves
the scaling regime and $\rho_{DE}^0$ is the present value of the
dark energy density.

In the evolution of dark energy velocity, we have to take into
account the momentum conservation equation given
by (\ref{cons}). This equation implies that the dark
energy velocity must be discontinuous at the transition points since
the equation of state  jumps at those times whereas  the quantity
$a^4(1+w_{DE})\rho_{DE}\vec{v}_{DE}$ is constant, being
$\rho_{DE}$ continuous. With this in mind, we get the following
evolution for dark energy velocity in the CCM frame:
\begin{eqnarray}
\vec{v}_{DE}=\left\{\begin{array}{c}
\vec{v}^{early}_{DE}
\;\;\;\;\;\;\;\;\;\;\;\;\;\;\;\;\;\;\;\;\;\;\;\;\;\;\;\;\;\;\;\;a<a_{eq}\\
\\\frac{4}{3}a_{eq}a^{-1}\vec{v}^{early}_{DE}\;\;\;\;\;\;\;\;\;\;\;
a_{eq}<a<a_{T}
\\\\
\frac{4a_{eq}a_{T}^{-3w_{DE}}}{3(1+w_{DE})}a^{3w_{DE}-1}\vec{v}^{early}_{DE}
\;\;\;\;\;\;a>a_{T}\end{array}
\right. .\label{vsm}
\end{eqnarray}

\begin{figure}[h]
\vspace{1cm}
\begin{center}
{\epsfxsize=10cm\epsfbox{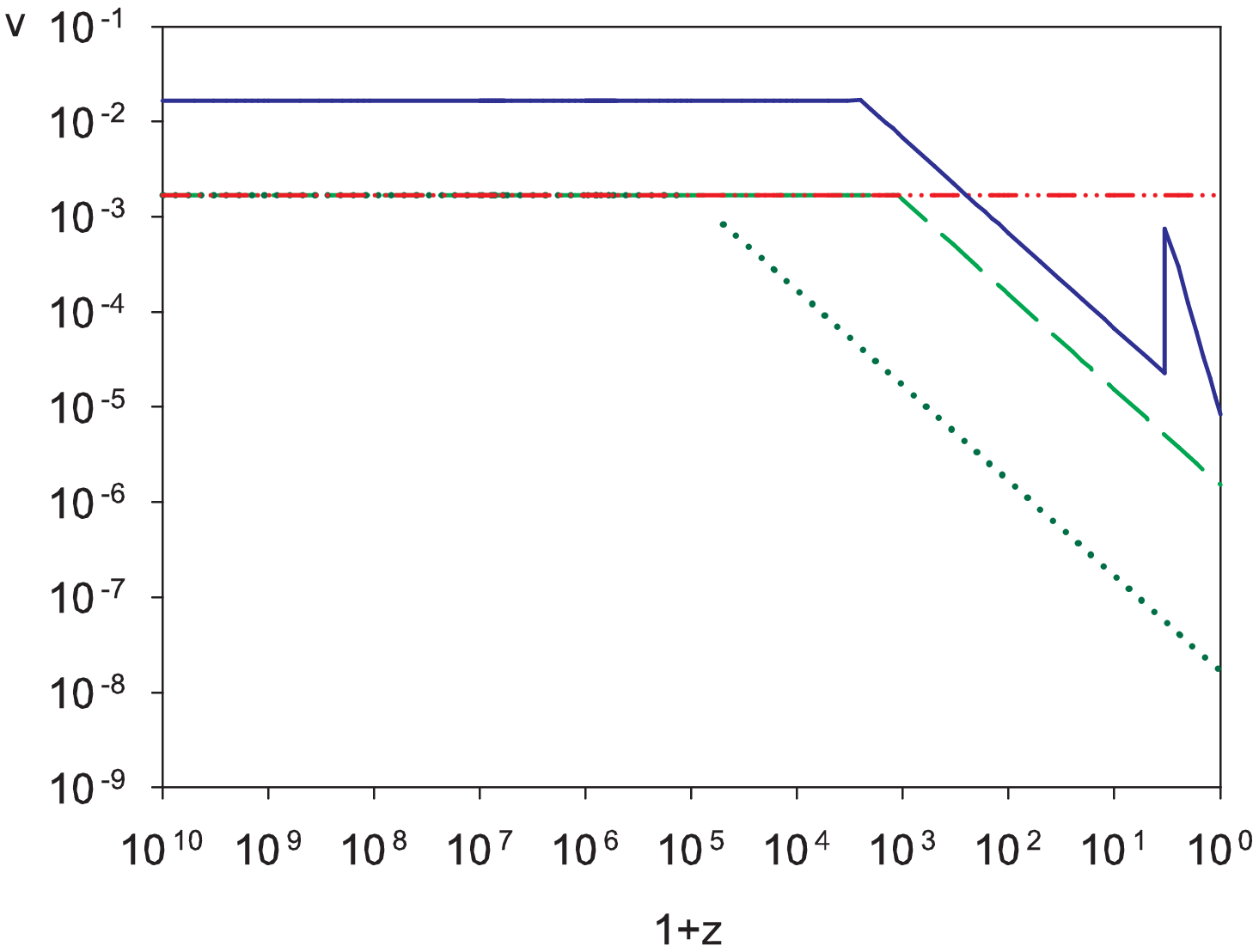}} \end{center}
{\footnotesize {\bf Figure 1:} 
Velocities evolution (CCM frame) in a scaling model with
$v_R=1.67\times10^{-3}\,$c and $\epsilon=0.1$. The continuous line
(blue) is for dark energy, dashed-dotted (red) for radiation,
dotted (cyan) for dark matter and dashed (green) for baryonic
matter. Notice that in this plot dark matter is assumed to
decouple at $z_*\simeq 10^5$. \cite{quadrupole}}
\end{figure}

In Fig. 1, we see the typical behavior of fluids velocities.
In these models, the quadrupole can be
approximated by (see \cite{quadrupole})
$Q_A\simeq 0.44\,\epsilon \,(\vec{v}_{DE}^{early})^2$
where $\vec{v}_{DE}^{early}$ relates to radiation velocity through
(\ref{velrel}) as follows: $v_R=\epsilon\, v_{DE}^{early}$ being
$\epsilon$ the initial fraction of dark energy density.
 With this
relation we can express the quadrupole as $Q_A\simeq 0.44\,
v_R^2\,\epsilon^{-1}\simeq1.23\times10^{-6}\epsilon^{-1}$ or,
equivalently, we have that $h\simeq
5.32\times10^{-6}\epsilon^{-1}$, where we have assumed that the
total bulk flow is due to the motion of the fluids and we have taken the
velocity of radiation to be $v_R\simeq 500$ km/s $=1.67\times
10^{-3}\,$c in the CCM rest frame. Therefore, constraints
(\ref{hexcluded}) read:
\begin{eqnarray}
7.69\times10^{-2}\;\lsim &\,\epsilon\,&\lsim\, 0.2 \;\;\;\;\;\ 68\% \;\mbox {C.L.}\nonumber \\
4.32\times10^{-2}\;\lsim &\,\epsilon\,&\lsim\, 0.2 \;\;\;\;\;\ 95\%
\;\mbox {C.L.}
\end{eqnarray}
where the upper limit comes from primordial nucleosynthesis, which
imposes the amount of dark energy density at that time to be less
than about 20$\%$ \cite{Kolb} so that $\epsilon\lsim \,0.2$.

To summarize,  in this kind of models it is possible to explain the
presence of the matter bulk flow from the dark energy motion in a
compatible way with the measurements of the CMB quadrupole.

\section{Discussion and conclusions}

Finally, we would like to comment on the fact that the total
quadrupole has a preferred axis which happens to coincide with
the direction of the velocities and, as a consequence, with that
of the dipole. Therefore, a moving dark energy model could also
shed some light on the so-called axis of evil problem \cite{AOE}.
Although this anomaly usually refers to the observed alignment of
the $\ell=2-5$ multipoles, there are also evidence that the axis
of such alignment is correlated with the dipole direction at more
than $99\%$ C.L. \cite{Copi}. Since moving dark energy gives a
common physical mechanism for both the dipole and quadrupole
contributions, it is expected to have correlations among them.
Indeed, in
\cite{quadrupole} it is shown that the motion of dark energy could
solve the low quadrupole anomaly for some models of dark energy,
in particular, for scaling and null dark energy models. The
solution of the low quadrupole problem arises because the relative
motion of the fluids generates a certain degree of anisotropy
which is seen by the photons coming from the last scattering
surface and acquire a quadrupolar anisotropy.  In a frame with the
$z$-axis pointing along the fluids motions, the power of the
quadrupole given by the generated anisotropy is zonal, i.e., it is
concentrated in the $m=0$ component so that it gives rise to a
cylindrical contribution. However, we still have to add the
standard isotropic fluctuation generated during inflation, whose
components are all comparable. Then, if we add linearly both
contributions in such a way that the resulting quadrupole is lower
than the inflationary one, the suppression has to take place for the
$m=0$ component and, therefore, the total quadrupole will be
non-cylindrical.

Thus, if we want to explain the low quadrupole with
this new contribution plus the standard inflation contribution, we need
\cite{quadrupole}:
\begin{eqnarray}
54\;\mu\K^2\lsim\;&(\delta T_A)^2&\;\lsim
3857\;\mu\K^2 \;\;\ 68\% \;\mbox {C.L.}\nonumber \\
0\;\mu\K^2\lsim\;&(\delta T_A)^2&\;\lsim 9256\;\mu\K^2 \;\;\ 95\%
\;\mbox {C.L.} \label{constraint}
\end{eqnarray}
which leads to the following constraints on $h$:
\begin{eqnarray}
1.18\times10^{-5}\;\lsim\; &h& \;\lsim\; 9.96\times10^{-5}
\;\;\ 68\% \;\mbox {C.L.}\nonumber \\
0\;\lsim\; &h&\; \lsim\; 1.54\times10^{-4} \;\;\ 95\% \;\mbox {C.L.}
\label{hconstraint}
\end{eqnarray}

For scaling models,  we have that the low
quadrupole can be explained if the following constraints hold:
\begin{eqnarray}
5.34\times10^{-2}\;\lsim \;&\epsilon&\;\lsim 0.2 \;\;\;\;\;\ 68\%
\;\mbox {C.L.}\nonumber \\
3.45\times10^{-2}\;\lsim \;&\epsilon&\;\lsim 0.2 \;\;\;\;\;\ 95\%
\;\mbox {C.L.}
\end{eqnarray}

\begin{figure}\begin{center}
{\epsfxsize=10.0 cm\epsfbox{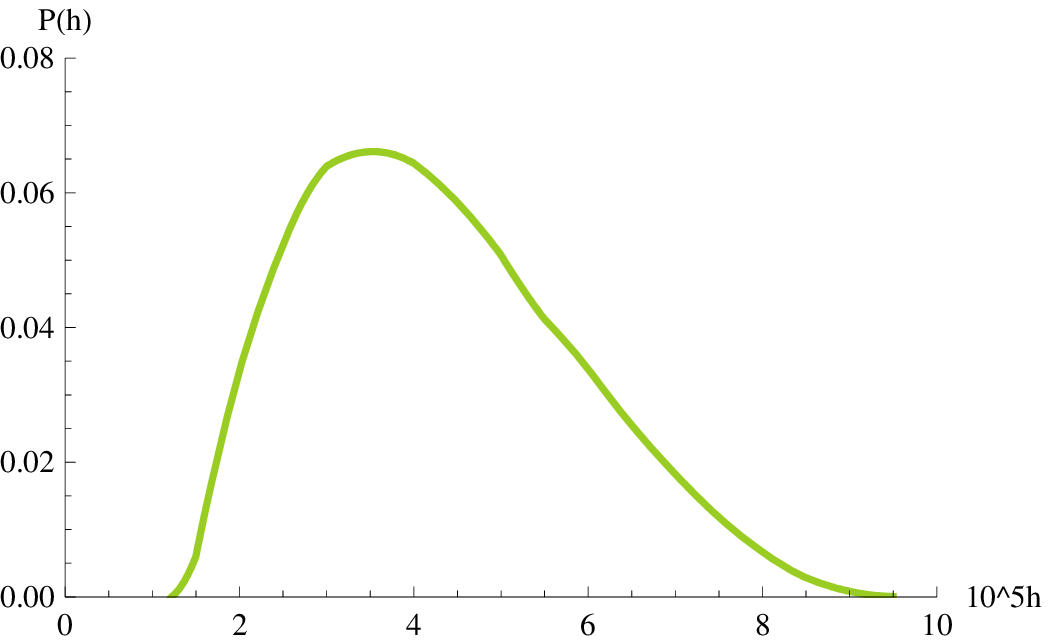}}
\end{center} 
{\footnotesize {\bf Figure 2:} Probability 
of having a quadrupole within 
$1\sigma$ interval of the measurements as a function of the parameter
$h$, which contains the information about the dark energy model.
}\label{fig}
\end{figure}

As the initial conditions are supposed to be random, we can
compute the likelihood of explaining the low quadrupole with the
anisotropic contribution as a function of the model-dependent
parameter $h$. To that end, we perform simulations of the total
quadrupole by choosing the phases $\alpha_i$ and the direction
$(\theta,\phi)$ randomly. In Fig. 2 we can see that the maximum of
the likelihood happens for $h\simeq 3.5\times 10^{-5}$ for which
the chance of having a quadrupole as low as the observed one is
$\simeq 7\%$.

In conclusion, we have shown that the detected large scale matter bulk flow
could be signalling the presence of a dark energy flow which would be
responsible for the present relative motion of matter and radiation.
This kind of explanation could also shed some light on the problem
of the low quadrupole and the alignment of dipole and quadrupole axes.

\vspace{0.1cm}

{\em Acknowledgments:}
We would like to thank Fernando Atrio Barandela for 
useful discussions and suggestions. 
This work has been  supported by
DGICYT (Spain) project numbers FPA 2004-02602 and FPA
2005-02327, UCM-Santander PR34/07-15875, CAM/UCM 910309 and
MEC grant BES-2006-12059.

\end{document}